\documentclass[10.5pt,twocolumn,showpacs,showkeys,preprintnumbers,amsmath,amssymb,prb,superscriptaddress]{revtex4-2}
\usepackage[utf8]{inputenc}
\usepackage{amsmath}
\usepackage{amsfonts}
\usepackage{epsfig}
\usepackage{bm}
\usepackage{setspace}
\usepackage{booktabs} 
\usepackage{graphicx}
\usepackage[colorlinks=true, citecolor=red, linkcolor=blue, urlcolor=blue]{hyperref}
\usepackage{float}
\usepackage[section]{placeins}
\usepackage{siunitx}
\usepackage{multirow}
\usepackage{array}
\usepackage{placeins}
\usepackage{afterpage}
\newcolumntype{C}[1]{>{\centering\arraybackslash}m{#1}} 
\begin{document}
\title{Optical spin injection in graphane and fluorographene}
\author{Ang\'elica Marina L\'opez-Mart\'inez}
\affiliation{Universidad de Ciencias y Artes de Chiapas. Instituto de Investigaci\'on e Innovaci\'on en Energ\'ias Renovables, libramiento norte poniente 1150 Lajas Maciel, 29039 Tuxtla Guti\'errez, Chiapas, M\'exico}
\author{C\'esar Castillo-Quevedo}
\affiliation{Departamento de Ciencias B\'asicas, Centro Universitario de la Ci\'enega, Universidad de Guadalajara, Av. Universidad, N\'um.1115, Col. Lindavista, C\'odigo Postal 47810, Ocotl\'an, Jalisco, M\'exico.}
\author{Cesar Camas-Flores}
\affiliation{Ingenier\'ia en Nanotecnolog\'ia, Universidad Polit\'ecnica de Chiapas, Carretera Tuxtla Guti\'errez-Portillo Zaragoza km 21+500, Col. Las Brisas, Suchiapa 29150, Chiapas, M\'exico.}
\author{Analila Luna-Valenzuela}
\affiliation{Universidad Aut\'onoma de Occidente (UAdeO), Unidad Regional Los Mochis. Departamento Acad\'emico de Ciencias de la Salud,  Blvd. Macario Gaxiola y Carretera Internacional, M\'exico 15, C.P. 81223, Los Mochis, Sinaloa, M\'exico.}
\author{Jose Luis Cabellos}\email[email:]{jose.cabellos@uptapachula.edu.mx}
\affiliation{Coordinaci\'on Ingenier\'ia en Mecatr\'onica, Universidad Polit\'ecnica de Tapachula, Carretera Tapachula a Puerto Madero km 24+300, San Benito, Puerto Madero C.P. 30830, Tapachula, Chiapas, M\'exico}
\author{Heber Vilchis-Bravo}
\affiliation{Universidad de Ciencias y Artes de Chiapas. Instituto de Investigaci\'on e Innovaci\'on en Energ\'ias Renovables, libramiento norte poniente 1150 Lajas Maciel, 29039 Tuxtla Guti\'errez, Chiapas, M\'exico}
\date{\today}
\begin{abstract}
  We theoretically investigate the optical spin-injection response in different stoichiometric configurations of graphane and fluorographene using density functional theory. Our goal is to determine which configuration yields the strongest degree of spin polarization. The results show that the fluorographene zigzag configuration yields the best  degree of spin polarization response (${\cal DSP}^{\mathrm{z}}$), with 98\% spin-polarized electrons at the band edge and  over a wide range of excitation photon energies.  In contrast, other graphane and fluorographene configurations achieve a ${\cal DSP}^{\mathrm{z}}$  of roughly 83–100\%, but only within a limited photon-excitation energy range.  In structures with low spin-orbit coupling, the degree of spin polarization is close to 100\% over a wide range of photon energies. For higher spin-orbit coupling, this strong response appears, but only in a narrow photon energy region. Additionally, under the band-resolved decomposition scheme, the contributions of different band-to-band transitions to the ${\cal DSP}^{\mathrm{z}}$ spectrum are identified by summing only the selected valence and conduction bands. Our findings show that almost the entire ${\cal DSP}^{\mathrm{z}}$ spectrum of the fluorographene zigzag configuration comes from transitions that involve only the top valence band, which is a mixture of C-p and F-p states.
\end{abstract}
\maketitle
\section{Introduction}
Spin injection in non-magnetic materials has become a key area of research in spintronics due to its potential for developing efficient, low-power devices~\cite{doi:10.1126/science.1065389,RevModPhys.87.1213,Miron2011}. Spintronics is an emerging multidisciplinary field~\cite{HIROHATA2020166711,RevModPhys.76.323}  that aims to understand the mechanisms of electrical and optical spin injection, the generation of optical spin orientation, and the detection of spin polarization. It also focuses on the propagation, transport, control, and detection of spin-polarized currents, leveraging this knowledge to develop functional devices~\cite{doi:10.1126/science.1065389,RevModPhys.76.323,Guo2024,PhysRevB.85.165324,PhysRevB.76.205113}. 
Since the 1990s, a variety of spin-based devices have been proposed; typical examples include spin-based transistors~\cite{10.1063/1.102730,JOHNSON1996321,10.1063/5.0276047,Taniyama2011,MI2023100408}, spin light-emitting diodes~\cite{Chang2025,Fiederling1999}, and spin-transfer-torque memories~\cite{KATINE20081217}. Electron spin is an intrinsic quantum property of an electron~\cite{10.1063/5.0133335}, first observed in the 1920s~\cite{Castelvecchi2022} through the Stern--Gerlach experiment.  It can have two orientations referred to as spin-up and spin-down~\cite{doi:10.1126/science.1065389}.  In early 1968,  Lampel~\cite{PhysRevLett.20.491} demonstrated the optical spin orientation, and  Pierce and Meier~\cite{PhysRevB.13.5484} subsequently developed an approach for optically detecting spin-oriented electrons emitted from GaAs using a Mott detector. Given this, opto--spintronics has drawn much attention because it can create spin currents and a non‐equilibrium spin population by absorption of circularly polarized light that transfers angular momentum without the need for direct contact with magnetic material~\cite{MIAH2011252, PhysRevB.90.035210,McIver2012,PhysRevLett.20.491}. This ultrafast mechanism is central to spintronics, with theoretical or experimental demonstrations in systems such as graphene~\cite{,doi:10.1021/acsnano.7b06800,https://doi.org/10.1002/pssb.201552565,PhysRevB.90.035210}, III–V semiconductors~\cite{10.1063/1.123515,PhysRevLett.20.491}, and 2D   transition-metal dichalcogenides (TMDC) heterostructures~\cite{Mak2012,Zeng2012,RevModPhys.90.021001,PhysRevB.92.155403}. Among the most studied materials, graphene has been recognized for its exceptional electronic properties such as high electronic charge mobility, weak spin-orbit coupling, negligible hyperfine interaction, and gate tunability, which are properties with potential applications  in optoelectronic and spintronic devices~\cite{RevModPhys.92.021003}. Previous studies have presented the experimental and theoretical results of spin injection in graphene, and have identified the advantages of graphene for spintronics compared to metals and semiconductors~\cite{Han2014}, and  challenges of graphene-based spintronics~\cite{RevModPhys.92.021003,10.1063/5.0191362}. However, one of the main challenges in using pristine graphene for spintronics is its weak spin-orbit coupling (SOC) and low optical absorption~\cite{doi:10.1021/acsnano.7b06800}, which hinders efficient spin injection and detection. To overcome this limitation, graphene hydrogenation has been explored as a functionalization technique capable of modifying its optical and spintronic properties. The addition of hydrogen and fluorine atoms to the graphene structure displaces carbon atoms out of the plane, altering the carbon-carbon bond length, which results in a band gap opening and a significant enhancement of spin-orbit coupling. This, in turn, improves graphene’s ability to transport and manipulate spin~\cite{Balakrishnan2014,doi:10.1021/nl802098g,doi:10.1126/science.1167130}. A similar behavior has been reported for fluorinated graphene, in which fluorine functionalization significantly enhances the spin--orbit interaction~\cite{PhysRevLett.108.226602}. Optical spin injection in functionalized graphene is a relatively new topic and has received less attention than electrical spin injection. However, relevant experimental studies using optical techniques such as {{Kerr}} spectroscopy and photoluminescence~\cite{Soriano_2015,Laterza:22,KHAMARI2019204} have demonstrated that graphene functionalization can effectively induce key optical and spintronic properties, such as enhanced spin-orbit interaction and the creation of localized spin centers. On the theoretical side, research has focused on understanding how graphene functionalization (primarily with H and F atoms) affects its electronic properties and spin-orbit coupling. Theoretical methods employed include first-principles calculations~\cite{PhysRevB.92.235444,PhysRevLett.110.246602} and simulations, showing that hydrogen functionalization and defects induce local magnetic moments in graphene~\cite{PhysRevB.75.125408,doi:10.1021/nl802234n, doi:10.1126/science.1167130}, which could be exploited for optical spin injection.
\begin{figure}[htb!]
\begin{center}  
  \includegraphics[scale=0.6]{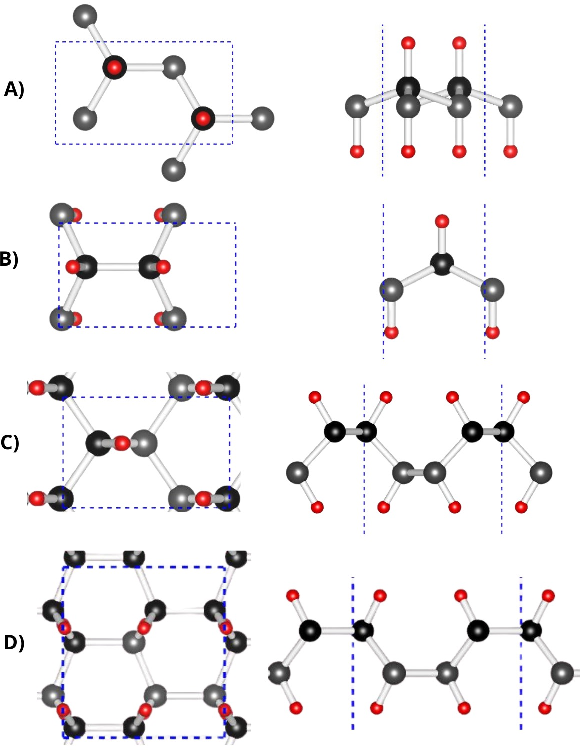}
  \caption{(Color online) The different configurations used in the calculations are displayed in two orientations (front and side views) of hydrogen/fluorine-graphene: A) chair, B) boat, C) zigzag, and D) armchair configurations.   The blue-dashed lines indicate the supercells used in the calculations. Note that all unit cells are rectangular. The black spheres represent carbon atoms, while the red spheres represent hydrogen or fluorine atoms.}
\label{opt}
\end{center}    
\end{figure}
In this study, we investigate the degree of spin
polarization along $\hat{{\bf{z}}}$  ({${\cal DSP}^{\mathrm{z}}$}) direction,   induced by circularly polarized light in fully functionalized graphene with
hydrogen (H) and fluorine (F) atoms. We consider four possible
geometric configurations: chair, boat, zigzag,
and armchair. These geometric configurations are based
on the study conducted by O. Leenaerts et al.~\cite{PhysRevB.82.195436}. The purpose of this work is to elucidate the
influence of adatom functionalization on the spin response of the studied materials and to identify which configuration most
effectively enhances optical spin injection. Nastos et al.~\cite{PhysRevB.76.205113} demonstrated that one can compute the
degree of spin polarization from full DFT band structures and refine it by applying a scissors correction to align the
theoretical band gap with the experimental value~\cite{PhysRevB.85.165324,PhysRevB.72.045223}. Therefore, in this work, we employed a comprehensive electronic band structure scheme within DFT to study optical spin injection under illumination with circularly polarized light  in graphane and fluorographene configurations. The results indicate that the fluorographene configuration (C) yields the best signal, achieving 98\% and -57\% spin polarization over a wide range of excitation photon energies. In comparison, the fluorographene configuration (A) reaches 100\% spin polarization, but only over a narrow window of photon energy.\\
\vspace{-0.5cm} 
\section{Theory}
In this section, we present the theoretical framework and the most important
expressions for calculating the spin injection and {${\cal DSP}^{\mathrm{z}}$}.  In this work, the calculations of optical spin injection and the carrier injection rate  in graphane and  fluorographene systems  are  described for an incident
monochromatic field of a frequency $\omega$ given  by Equation~\ref{max}
\begin{eqnarray}\label{max}
  {\bf{E}}(t)={\bf{E}}{(\omega)}e^{(-i\omega t)}+ 
  {\bf{E}}^{*}{(\omega)}e^{(i\omega t)}
\end{eqnarray}
The {${\cal DSP}^{\mathrm{z}}$}, which  quantifies the excess of 
spin-up polarized carriers ($\eta_{\uparrow}$) over the spin-down polarized carriers ($\eta_{\downarrow}$) is given by Equation~\ref{spincito}.  
\begin{equation}
  \displaystyle
      {\cal DSP}^{\mathrm{a}}= \frac{\eta_{\uparrow}-\eta_{\downarrow}}{\eta_{\uparrow}+\eta_{\downarrow}}
        \label{spincito}
\end{equation}
and the  {${\cal DSP}^{\mathrm{a}}$}  along  the $\hat{\bf{z}}$  direction is formally given in terms of the injection rates by Equation~\ref{eq:dsp}
\begin{eqnarray}\label{eq:dsp}
  {\cal DSP}^{\mathrm{a}} =\frac{\dot S^{\mathrm{a}}}{(\hbar/2)\dot n}, 
\end{eqnarray}
where $\dot n $   is the carrier injection rate, $\hbar $ is the Planck constant, $\dot n $   is given by
Equation~\ref{portadora} considering only the one-photon transition amplitude. 
The theory of {${\cal DSP}^{\mathrm{a}}$} is laid out by Nastos et al.~\cite{PhysRevB.76.205113}, and Sipe et al.~\cite{PhysRevB.61.5337}, where we refer the reader for the details. Here, we reproduce only the most important expressions to calculate the {${\cal DSP}^{\mathrm{z}}$}. 
Previous works employed a similar theory~\cite{PhysRevB.80.245204,PhysRevB.85.165324,https://doi.org/10.1002/pssb.201552565}. 
The $\dot n $   is given by
Equation~\ref{portadora} considering only the one-photon transition amplitude. 
\begin{equation}
     \displaystyle
     \dot n=\xi^{{ab}}(\omega) {\bf{E}}^a{(-\omega)}{\bf{E}}^{b}{(\omega)}.
     \label{portadora}
\end{equation}
with the  tensor $\xi$  given by Equation~\ref{chi} 
\begin{eqnarray}
     \label{chi} 
      \displaystyle
   \xi^{{ab}}(\omega)&=&\frac{2\pi e^2}{\hbar^2} \sum_{c,v} 
                            \int \frac{d^{3}k}{8\pi^3} 
                             r^a_{v,c} ({\bf{k}})
                             r^b_{c,v} ({\bf{k}})\\
\nonumber  
  & & \times (\delta[\omega_{cv}({\bf{k}})-\omega].
 \end{eqnarray} 
The $\xi^{ab}(\omega)$  response tensor is related to  $\chi^{ab}(\omega)$,  the imaginary part of the linear
optical response tensor~\cite{PhysRevB.76.205113}, by $4\pi \text{Im}[\chi^{ab}(\omega)]= 2\pi\hbar \xi^{ab}(\omega)$  in the Gaussian system.  
Similarly, the rate of one-photon spin injection is given by Equation~\ref{spino}
\begin{equation}
  \label{spino} 
     \displaystyle
\dot S^{\mathrm{a}}= \zeta^{\mathrm{abc}}(\omega)
E^{\mathrm{b}}(-\omega) E^{\mathrm{c}}(\omega) 
\end{equation}
The pseudotensor $  \zeta^{\mathrm{abc}}(\omega)$ is purely imaginary and is given by Equation~\ref{zetaabci}
\begin{eqnarray}\label{zetaabci}
\zeta^{\mathrm{abc}}(\omega)
&=&
\frac{i\pi e^2}{\hbar^2}
\int\frac{d^3k}{8\pi^3}
\sum_{vcc'}\,'\,
\mathrm{Im}\Big[S^{\mathrm{a}}_{c'c}(\mathbf{k}) r^{\mathrm{b}}_{vc'}(\mathbf{k}) r^{\mathrm{c}}_{cv}(\mathbf{k})
\nonumber\\
&+& 
S^{\mathrm{a}}_{cc'}(\mathbf{k}) r^{\mathrm{b}}_{vc}(\mathbf{k}) r^{\mathrm{c}}_{c'v}(\mathbf{k})\Big]
 \delta(\omega_{cv}(\mathbf{k})-\omega)
,
\end{eqnarray}  
A multiple-scale approach is employed to
consider the coherence between  degenerate excited 
states~\cite{PhysRevB.76.205113,PhysRevB.107.165202}. 
 The summation in Equation~\ref{zetaabci} is restricted to
the conduction bands  (c) and (c') that
are separated by less than 30 meV~\cite{PhysRevB.76.205113}.   The nonzero independent
component different from zero for the  $  \zeta^{\mathrm{abc}}(\omega)$  pseudotensor is
$ \zeta^{\mathrm{zxy}}$ for graphane and fluorographene systems. 
The matrix elements of the position operator
  $\hat {r}^{a}_{cv}$ and spin operator   $\hat {s}^{a}_{cv}$  
 and the
energy difference between the valence
(v) and conduction states, (c),  $\omega_{cv}({\bf{k}})$,   are
evaluated at {{\bf{k}}}-points on a specially designed
tetrahedral grid. This grid is used in the
integrals of Equations~\ref{chi} and~\ref{zetaabci}, which are
computed  using  an  analytical  linear
tetrahedral integration method over a {\bf{k}}-point
mesh~\cite{PhysRevB.72.045223, PhysRevB.80.245204}. 

\section{Computational details}
The calculations were conducted using the freely available ABINIT software package~\cite{Gonze2016,GONZE2002478,GONZE20092582,Gonze+2005+558+562}. In particular, we computed the real-space Bloch wavefunctions and the band structure. We generated the data required for the SPIN code developed by A. Marina L\'opez based on the TINIBA code developed  by B. Mendoza, J. L. Cabellos, and T. Rangel~\cite{doi:10.1021/acs.nanolett.4c03880,PhysRevMaterials.8.116203} which is capable of computing optical spin injection. With ABINIT, we carried out geometry optimization, relaxing the atomic positions until the maximum force per atom was less than  20 meV/\AA.
 We employed the Perdew–Burke–Ernzerhof functional (PBE)  within the generalized gradient approximation (GGA) in the framework of DFT~\cite{PhysRevLett.77.3865} to include the exchange correlation energy.  To account for the dispersion effects (van der Waals corrections), we employed  Grimme’s D3 correction~\cite{10.1063/1.3382344,Shi2019,doi:10.1021/acsomega.1c03849} as implemented in the ABINIT code. Dispersion interactions are important for accurately describing adsorption of water on graphene~\cite{PhysRevB.86.195436} or adatom-graphene geometries: neglecting vdW corrections leads to incorrect adsorption sites, and misleading trends across adsorbates~\cite{PhysRevB.85.205402}.  We employ a norm-conserving pseudopotential,  specifically the Hartwigsen-Goedecker-Hutter (HGH) pseudopotential~\cite{PhysRevB.58.3641},  incorporates scalar relativistic effects~\cite{Berry2024}. These pseudopotentials do not include non-linear core corrections and can be used to perform mGGA calculations~\cite{doi:10.1126/science.aad3000}.  The Bloch wavefunctions were expanded in a plane-wave basis set, and their convergence was verified using a kinetic cutoff of 30 Ha. The Monkhorst--Pack scheme was used to sample the irreducible Brillouin zone (IBZ)~\cite{PhysRevB.13.5188} with a 20x20x1 {\bf{k}}-point grid, and the integration over the IBZ was carried out using the
tetrahedron method~\cite{PhysRevB.80.245204,PhysRevB.85.165324}.
Contributions to the matrix elements from the non-local part of the
pseudopotential  are excluded~\cite{PhysRevB.44.13071}. We have neglected the effects of local fields
and excitonic effects, a theoretical challenge that should be addressed~\cite{RevModPhys.74.601}. We restrict our analysis to the instantaneous spin polarization generated by circularly polarized light. Electron–phonon thermalization, particularly due to optical phonons, is not included.  

\section{Results}
\subsection{Structural properties}
\begin{table}[htb!]
  \centering
  \begin{tabular}{lcccccccc}
    \toprule
      & \multicolumn{4}{c}{\textbf{Graphane}} 
      & \multicolumn{4}{c}{\textbf{Fluorographene}} \\
    \cmidrule(lr){2-5} \cmidrule(lr){6-9}
      & A  & B & C & D
      & A  & B & C & D \\ 
    \midrule
    a$_0$ (\textbf{\AA})   
      & 4.39  & 4.30 & 3.82 & 4.31
      & 4.50  & 4.51 & 4.18 & 4.61 \\
    b$_0$ (\textbf{\AA})     
      & 2.54 & 2.52 & 2.54 & 4.54
      & 2.60 & 2.52 & 2.62 & 4.89 \\
    \bottomrule
  \end{tabular}
   \caption{The equilibrium lattice constant a$_0$ for the different stoichiometric configurations was obtained by performing a full geometry optimization within DFT at the PBE-GGA GD3 dispersion correction~\cite{10.1063/1.3382344,Shi2019,doi:10.1021/acsomega.1c03849}  level of theory, including a vacuum region of 20~\AA~to avoid interactions between periodic images. A, B, C, and D are the configurations displayed in Figure~\ref{opt}.}
  \label{LCON}
\end{table}

 As mentioned previously, this study aims to
evaluate the {${\cal DSP}^{\mathrm{z}}$} of injected electrons in graphene functionalized with
hydrogen and fluorine atoms in four different
geometric configurations shown in the Figure~\ref{opt}. Table~\ref{LCON} presents a comparison of the calculated equilibrium lattice constants a$_0$ for four stable hydrogenated and fluorinated graphene configurations, which are schematically depicted in Figure~\ref{opt}. These configurations are referred to as the chair, boat, zigzag, and armchair conformations. For the chair graphane structure, the optimized lattice parameters are b=2.540 ~\AA~and a=4.399~\AA, in good agreement with previous theoretical results for graphane, where b=2.534~\AA~ and a=4.3769~were reported. We note a lattice expansion from 2.46~\AA~for pristine graphene~\cite{Meyer2007} to our calculated value of 2.540~\AA. This expansion arises from the change in carbon hybridization from sp$^2$ to sp$^3$ upon hydrogen adsorption on both sides of the graphene sheet. From the experimental point of view, Sofo et al. \cite{PhysRevB.75.153401} reported an optimized lattice constant of 2.51~\AA~ for fully hydrogenated graphene, consistent with our calculated values. For the boat and zigzag graphane configurations, the computed lattice constants are b=2.520~\AA,~a=4.300~\AA~and
 b=2.540~\AA~, a=3.820~\AA,~respectively, the values  are comparable to those of the chair conformation. The armchair graphane unit cell is nearly square, with  b=4.540~\AA~ and a=4.310~\AA,~ showing a noticeably different geometry from the other three structures. The four fluorographene conformations follow the same general trend as their hydrogenated counterparts, although their optimized lattice parameters differ slightly, as summarized in Table I. We emphasize that the distinct configurations exhibit noticeable structural distortions, indicating that the adsorption sites strongly modify the local symmetry and, consequently, the optical properties,  particularly those related to optical spin injection.
\begin{table}[htb!]
  \centering
 \resizebox{\columnwidth}{!}{\begin{tabular}{lcccccccc}
    \toprule
      & \multicolumn{4}{c}{\textbf{Graphane}} 
      & \multicolumn{4}{c}{\textbf{Fluorographene}} \\
    \cmidrule(lr){2-5} \cmidrule(lr){6-9}
      & A  & B & C & D
      & A  & B & C & D \\ 
    \midrule
    C-C ({\AA})   
      & 1.536  & 1.558 & 1.543 & 1.550
      & 1.558  & 1.580 & 1.586 & 1.604 \\
    C-H/F ({\AA})     
      & 1.111 & 1.104 & 1.105 & 1.102
     & 1.382 & 1.380 & 1.382 & 1.376 \\
    $\Delta_Z$ ({\AA}) 
      & 0.460 & 0.649 & 1.139 & 1.143
    & 0.487 & 0.625 & 1.020 & 1.110 \\
    $\theta$[CCC] $^{\circ}$
    & 111.5 & 112.4 & 112.4 & 112.4
    & 110.9 & 114.0 & 115.5 & 113.9 \\
    $\theta$[CC(H/F)] $^{\circ}$
    & 107.4 & 107.2 & 107.6 & 106.6
    & 108.0 & 107.8 & 103.6 & 106.5 \\
    \bottomrule
  \end{tabular} } 
 \caption{Structure parameters for the four different A, B, C, and D configurations
   displayed in Figure~\ref{opt}.  For each configuration we report the bond distance C--C, C--H  in~\AA,~ the out-of-plane  buckling height  $\Delta_Z$  in~\AA,~
   and the angle $\theta$[CCC] and the bond angles $\theta$[CCH] and  $\theta$[CCF]. }
  \label{angulos}
\end{table}

Table~\ref{angulos} presents the distance between
neighboring C atoms, dCC, and the angles,
$\theta$~CCX. In general, C--C bond lengths in graphane are
 estimated to range between 1.52--1.56~\AA~\cite{C3CS60132C}, which are larger than those reported for  pristine graphene of 1.42~\AA~\cite{RevModPhys.81.109}. This increase is attributed to the structural expansion  resulting from the incorporation of hydrogen or fluorine atoms that modify the sp$^2$ to sp$^3$.  The interatomic distances are slightly larger in fluorographene,  owing to the larger atomic radius of fluorine compared to that of hydrogen~\cite{doi:10.1126/science.1167130,https://doi.org/10.1002/smll.201001555,doi:10.1021/acs.chemrev.6b00664}. The average C--H bond dissociation energy is 413--415 kJ mol$^{-1}$, while C--F bond dissociation energies are in range from  439--485 kJ mol$^{-1}$~\cite{luo2002handbook}. However, there is a notable difference reported in previous works in their formation energies that indicate the fluorination is energetically more favorable than hydrogenation, with  values of about 0.9 eV per F atom for fluorographene~\cite{PhysRevB.83.115432} and only ~0.1 eV per H atom for graphane~\cite{PhysRevB.75.153401}, demonstrating the higher thermodynamic stability of C--F bonds compared to C--H bonds.

 \subsection{Band structure}
\begin{table}[htb!]
  \centering
   \resizebox{\columnwidth}{!}{\begin{tabular}{lllllllll}   
    \toprule
      & \multicolumn{4}{c}{\textbf{Graphane}} 
      & \multicolumn{4}{c}{\textbf{Fluorographene}} \\
    \cmidrule(lr){2-5} \cmidrule(lr){6-9}
    eV   & A  & B & C & D & A & B & C & D \\ 
    \midrule
    E$_{\textrm{gap}}$ 
      & 3.48  & 3.35 & 3.30 & 3.26
      & 2.74  & 3.05 & 3.28 & 4.27 \\
    E$_{\textrm{gap}}^{\mathrm{{GGA}}}$
      & 3.70 & 3.61 & 3.58 & 3.61
     & 3.20 & 3.23  & 3.59 & 4.23 \\
    E$_{\textrm{gap}}^{{GW}}$ 
      & 6.05 & 5.71 & 5.75 & 5.78
    &  7.42 & 7.32 & 7.28 & 7.98 \\
     E$_{\textrm{gap}}$ (Exp.)
    & 5.75 & -- & -- & --
    &3.8 & -- & -- & --- \\
    $\Delta_{\textrm{SO}} (1\times 10^{-6})$
    & 2420 & 1080 & 510 & 1400
    & 1720. &  60  &  380  &   1820 \\
    \bottomrule
   \end{tabular} }
    \caption{Our computed electronic band gaps PBE--GGA (E$_{\textrm{gap}}$) together with reference DFT--GGA and $GW$ band gaps (in eV) for fully hydrogenated graphene 
(graphane) in the A) chair, B) boat, C) zigzag, and D) armchair conformations 
shown in Figure~\ref{opt}. The values $E_{\textrm{gap}}^{\mathrm{GGA}}$ and 
$E_{\textrm{gap}}^{GW}$ for graphane are taken from 
Leenaerts~\textit{et al.}~\cite{PhysRevB.82.195436}. 
The experimental  band gap (E$_{\textrm{gap}}$ (Exp.)) of pure fluorographene~\cite{HRUBY2022152839} and the 
computed SOC band splitting ($\Delta_{\textrm{SO}}$, reported in units of $\mu$eV)
are included in the last two rows, respectively.}
  \label{table_so2}
\end{table}
The electronic band structure determines the optical properties of materials, such as optical absorption, reflection, and emission~\cite{PhysRevB.34.5390, RevModPhys.74.601,ma13194300, PhysRevB.80.115205}.  One of the main challenges in theoretical studies of energy bands in graphane using DFT is the underestimation of the band gap. At first glance, this underestimation arises because DFT is based on the ground state, while the electronic band gap is an excited-state property~\cite{RevModPhys.74.601}. Several correction methods have been developed to address this limitation. One of them is the scissors correction, which involves rigidly shifting the conduction bands by a specific energy value to match the experimental band gap~\cite {PhysRevLett.63.1719,PhysRevB.80.155205, PhysRevB.72.045223}. The GW approximation is another method that corrects the electronic band gaps in semiconductors by including many-body effects~\cite{PhysRev.139.A796,RevModPhys.74.601}. Table~\ref{table_so2} compares the electronic band-gap values obtained in this study with those reported in previous research.  The data in Table~\ref{table_so2} show that the hydrogenated graphene structures A and B, displayed in Figure~\ref {opt},  have direct band gaps of 3.48 eV and 3.35 eV at the $\Gamma$ point. We found these values using the PBE-GGA method for the chair and boat conformations.   In contrast, previous studies have reported direct band gap values of 3.70 eV and 3.61 eV at the $\Gamma$ point for the chair and boat conformations, respectively~\cite{PhysRevB.82.195436}. Other published work reported a direct band gap of 3.5 eV for the chair conformer and 3.7 eV for the boat conformer at the $\Gamma$ point~\cite{PhysRevB.75.153401}.
\begin{figure}[ht!]
\begin{center}                               
  \includegraphics[scale=0.55]{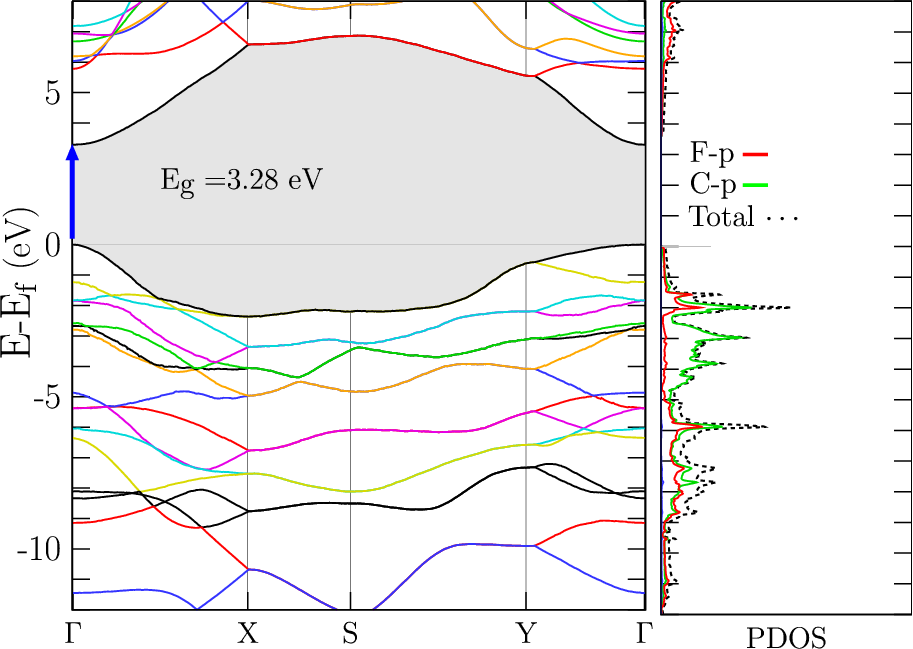}  
  \caption{(Color online) The left panel shows the calculated electronic band structure, while the right panel displays the partial density of states for the fluorographene configuration (C). We set the Fermi level to 0 eV to align the bands. The shaded region corresponds to the forbidden band, and the blue arrow indicates the band gap at the $\Gamma$ point of the Brillouin zone. The band structure is plotted without a scissors correction.}
\label{pdos}
\end{center}    
\end{figure}
The zigzag and armchair graphane conformations C and D exhibit band gaps of 3.30 eV and 3.26 eV, respectively. It is noteworthy that our calculated band gaps for graphane are slightly smaller than those reported by Leenaerts~\textit{et al.}~\cite{PhysRevB.82.195436}. We attribute this trend to dispersion corrections that primarily reduce the computed band gap by inducing small geometry-dependent shifts, an effect reported in dispersion-corrected DFT studies of solids, 2D materials, and molecules~\cite{PhysRevLett.102.073005, LUNAVALENZUELA2021102024}. There is a notable discrepancy between the experimental data for the chair hydrogenated configuration (A), which shows a band gap of 5.75 eV, and our PBE-GGA calculations. However, previous work reported the GW band-gap calculation that yields a value slightly higher by {5\%} than the experimental value. Furthermore, it's important to highlight that the trend of the band gap reported at the GW level by Leenaerts et al. \cite{PhysRevB.82.195436} for the four configurations is approximately twice that of our PBE-GGA band gap calculations across all geometric configurations. Experimental band gap values for fluorographene range from 3.1 to 3.8 eV,  indicating a wide band gap of greater than 3.0 eV~\cite{CHRONOPOULOS201760, https://doi.org/10.1002/smll.201001555},  while the theoretical value reported for the band gap at the GW level of theory is 7.42 eV for chair configuration~\cite{PhysRevB.82.195436}.  Our calculation using the PBE-GGA approach yields band gaps between 2.74 eV and 4.27 eV, as displayed in Table~\ref{table_so2}. Notice that these values are generally slightly smaller than those reported by Leenaerts~\textit{et al.}~\cite{PhysRevB.82.195436}  at the DFT-GGA level of theory. We attribute this trend to dispersion corrections. From the experimental point of view,  we consider that the experimentally measured band gaps are influenced by several factors, including the specific synthesis conditions, the degree of functionalization, and the measurement techniques employed~\cite{C5NR03243A}. Both (H) and (F) atoms adsorbed on graphene convert this from a sp$^2$ to an sp$^3$ system; However, the resulting hybridization differs among them. The average value of band gaps for the four configurations of hydrogenated graphene,  displayed in Table~\ref{table_so2},   is larger than that of fluorinated graphene because not just fluorinated graphene is closer to ideal sp$^3$ hybridization; In this work, we consider that a fluorine atom adsorbed on graphene introduces additional ionicity, which reduces the electronic band gap.
\begin{figure*}[hbt!]
\begin{center}                               
  \includegraphics[scale=0.95]{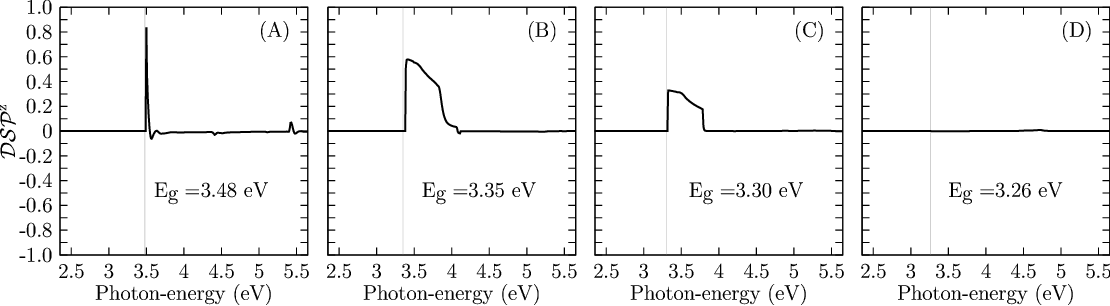}  
  \caption{(Color online) The degree of spin polarization of electrons,  {${\cal DSP}^{\mathrm{z}}$}, which is optically injected by one-photon absorption, is presented as a function of the photon energy in graphane under different configurations: A) chair, B) boat, C) zigzag, and D) armchair conﬁgurations. The light injects electrons that exhibit spin polarization in the z direction. The band-edge values of the {${\cal DSP}^{\mathrm{z}}$} of injected electrons are 83{\%},  58{\%},  33{\%}, and 0{\%}  for chair,  boat,  zigzag, and armchair conﬁgurations, respectively. E$_\textrm{g}$ is the {{Kohn-Sham}} band gap without scissor correction.}
\label{spinor1}
\end{center}    
\end{figure*}

\subsection{Partial density of states}
For deeper insight, we compute the partial density of states (PDOS) for fluorographene configuration (C) to identify which valence and conduction bands, and their orbital character, contribute to the {${\cal DSP}^{\mathrm{z}}$}. The left panel of Figure~\ref{pdos} displays the band structure, and the right panel its corresponding PDOS. The uppermost valence band,  situated near the Fermi level, in the energy range  0  to -2 eV, primarily consists of the hybridization of F--p and C--p states. In the band structure displayed on the left side of Figure~\ref{pdos}, the valley along Y--$\Gamma$--X appears closest to the Fermi level, indicating that it dominates the onset of the {${\cal DSP}^{\mathrm{z}}$} signal. The first conduction band is located around 3.30 eV above the Fermi level and shows a relatively parabolic dispersion along the  Y--$\Gamma$--X  direction. It is predominantly composed of a mixture of C--p and F--p states.

\subsection{Spin orbit coupling}
The computed SOC is presented in Table~\ref{table_so2} for hydrogenated and fluorinated graphene configurations. Previous studies have reported the SOC in pristine graphene to be 24 $\mu$eV~\cite{PhysRevLett.110.246602}, also employing the tight-binding theory of SOC in graphene Konschuh et al. report 24 $\mu$eV~\cite{PhysRevB.82.245412}. The SOC in pristine graphene is small because carbon is a light element with weak atomic spin-orbit interaction~\cite{PhysRevB.75.041401} and the planar sp$^2$ bonding further suppresses $\pi-\sigma$ hybridization~\cite{PhysRevLett.95.226801}. The Hydrogenated graphene configuration  (A) displayed in Figure~\ref{opt} exhibits the highest  computed value of SOC, with a value of 2420~$\mu$eV, and the smaller value of  SOC, with  60~$\mu$eV is exhibited for the fluorinated graphene configuration (B)
displayed in Figure~\ref{opt}. Hydrogen adsorption induces a large enhancement of SOC in graphene due to the formation of sp$^3$ hybridization, which increases the SOC  by one to two orders of magnitude compared to pristine graphene. The absorption of fluorine atom on graphene induces somewhat smaller SOC value as shown in Table~\ref{table_so2}. According to Table~\ref{angulos}, the average buckling,  $\Delta_Z$ , for the four hydrogenated graphene configurations,  $\Delta_Z$ =0.8477~{\AA},~ slightly larger than the average buckling of  0.8105~{\AA}~for the four fluorinated graphene configurations,  which implies that hydrogenated graphene has stronger sp$^3$ hybridization than fluorinated graphene. Fluorine atom is a heavier element and more electronegative than hydrogen atom. Because the fluorine is a heavier element, one might expect larger SOC; however, our results presented in Table~\ref{table_so2} indicate that the fluorine adsorbed on graphene induces less SOC than hydrogen adsorbed on graphene. The C-F bond is not purely covalent and has significant ionic character due to the fluorine atom being highly electronegative. The general reason for the enhancement of SOC is sp$^3$ hybridization, and  SOC is central for opto--spintronics/spintronics  in spin relaxation, spin transport,~\cite{PhysRevLett.110.246602}. Moreover,   large SOC is important for the spin Hall effect~\cite{PhysRevLett.110.156602,PhysRevLett.110.246602}.

\subsection{Optical spin injection in graphane}
\begin{figure*}[htb!]
\begin{center}  
  \includegraphics[scale=0.95]{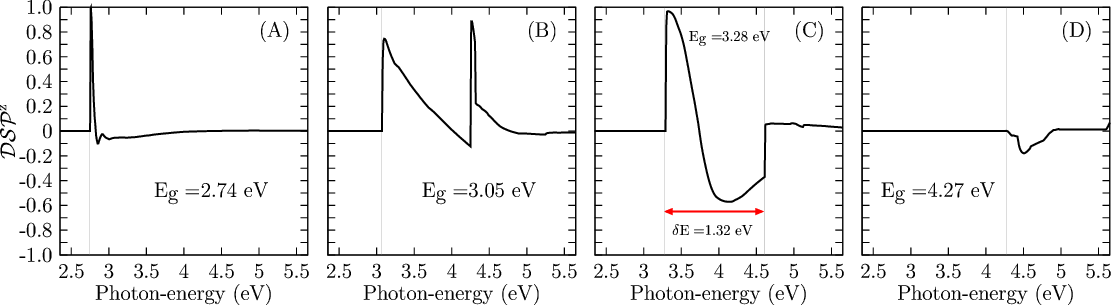}
  \caption{(Color online) The  {${\cal DSP}^{\mathrm{z}}$}, which is optically injected by one-photon absorption, is presented as a function of the photon energy in fluorographene under different configurations: A) chair, B) boat, C) zigzag, and D) armchair conﬁgurations. The light injects electrons that exhibit spin polarization in the z direction. The band-edge values of the  {${\cal DSP}^{\mathrm{z}}$} of injected electrons are 100{\%},  74{\%},  98{\%}, and -17{\%}  for chair,  boat,  zigzag, and armchair conﬁgurations, respectively. E$_\textrm{g}$ is the Kohn--Sham band gap without scissors correction. Figure 4(C),  $\delta$E=1.32 eV,  is the photo-excitation energy where   {${\cal DSP}^{\mathrm{z}}$} differs from zero marked with the red arrow. }
\label{spinor2}
\end{center}    
\end{figure*}

The results of the calculations for the ${\cal DSP}^{\mathrm{z}}$ of electrons in  $\hat{{\bf{z}}}$ direction are shown in Figure~\ref{spinor1} for the four graphane configurations --(A) chair, (B) boat, (C) zigzag, and (D) armchair-- as they are depicted in Figure~\ref{opt}. 
For circularly polarized light propagating along the  ${\hat{\bf{z}}} $ direction, the spin polarization of the injected electrons is aligned to the ${\hat{\bf{z}}}$-axis. To facilitate comparison among all the $\cal DSP^{\mathrm{z}}$ spectra shown in Figure~\ref{spinor1} and Figure~\ref{spinor2}, they are presented on the same scale for both the vertical $\cal DSP^{\mathrm{z}}$ axis and the  horizontal energy axis.
Moreover, in each of the four cases, the onset of the ${\cal DSP}^{\mathrm{z}}$  signal appears when the incident photon energy  ($\hbar\omega$) equals the DFT-calculated band gap. The spectra shown in the Figure~\ref{spinor1} and Figure~\ref{spinor2} correspond to unscissored  ${\cal DSP}^{\mathrm{z}}$  spectra. Notably,  applying the scissor correction does not modify the overall profile of either the  $\Im [\chi_1^{ab}(-\omega;\omega)]$ or the $\cal DSP^{\mathrm{z}}$ signals;  only their onset energies are shifted~\cite{PhysRevB.80.155205,PhysRevB.80.245204,PhysRevB.85.165324}. Panel (A) in Figure~\ref{spinor1},  shows the  $\cal DSP^{\mathrm{z}}$ spectrum for graphane configuration (A). The onset of the $\cal DSP^{\mathrm{z}}$ signal occurs at 3.48 eV on the horizontal energy axis. In Figure~\ref{spinor1}(A), we observe a narrow and intense $\cal DSP^{\mathrm{z}}$ peak that increases sharply in intensity. This $\cal DSP^{\mathrm{z}}$ peak has the appearance of a unit impulse function, but it has a finite excitation energy window of approximately 0.05 eV. Importantly,  the optically injected electrons achieve a ${\cal DSP}^{\mathrm{z}}$  of 83{\%}. However, the signal also decreases abruptly and vanishes rapidly just above the absorption onset, at approximately 3.53 eV. This behavior indicates that, in the geometric graphane configuration (A), the energy range in which spin polarization can be nonzero is very narrow, spanning only 0.05 eV. At energy of 3.53 eV, the  $\cal DSP^{\mathrm{z}}$ signal exhibits a small peak that takes a roughly negative value of 5\%. Beyond this energy region, the $\cal DSP^{\mathrm{z}}$ signal approaches zero. 
From a theoretical standpoint, our findings for the $\cal DSP^{\mathrm{z}}$, shown in Figure~\ref{spinor1}(A), are in agreement with previous reports of   $\cal DSP^{\mathrm{z}}$ in graphane-like structures with different hydrogen coverages on graphene. Zapata-Pe\~na et al. reported  values of 39\% and 57\% of  the  $\cal DSP^{\mathrm{z}}$  in C$_{16}$H$_8$ structures, with a narrow excitation-energy window of 0.002 eV~\cite{https://doi.org/10.1002/pssb.201552565}. Inglot et al. developed a theoretical model of optical spin injection in graphene with {{Rashba SOC}} and provided numerical predictions for the injection rates~\cite{PhysRevB.89.155411}. Avsar et al. reported experimental evidence of optical spin injection in graphene~\cite{doi:10.1021/acsnano.7b06800}. The graphane configuration (A) yields an 83\%  of   $\cal DSP^{\mathrm{z}}$ depicted in Figure~\ref{spinor1}(A). However, the excitation energy window is only 0.05 eV, which makes the graphane configuration (A) unsuitable for spintronic experimental implementation or device spintronic applications.

Figure~\ref{spinor1}(B) shows the  $\cal DSP^{\mathrm{z}}$ spectrum for the graphane configuration (B).  In Figure~\ref{opt}(B), we show the graphane configuration (B). At the onset of the $\cal DSP^{\mathrm{z}}$, which occurs at 3.35 eV, injected electrons are {58\%} spin-polarized. As the photon energy is increased above 3.35 eV, the   $\cal DSP^{\mathrm{z}}$ of electrons drops almost linearly. At a photon energy value of approximately 3.8 eV, the  $\cal DSP^{\mathrm{z}}$ of electrons reached the value of {35\%}, and at this point, it drops abruptly, almost vertically, to zero.  At a photon energy of 4.05 eV, the   $\cal DSP^{\mathrm{z}}$ becomes negative, with a value of 1\%. As  energy increases beyond 4.10 eV, the   $\cal DSP^{\mathrm{z}}$  is zero. The excitation-energy range relevant for spin injection spans roughly 0.70 eV. A comparable trend has been reported in calculations of $\mathcal{DSP}^{z}$ for GaAs and Si under tensile stress~\cite{PhysRevB.80.245204}. Owing to this wider energy window, graphane configuration (B) is a more favorable candidate for experimental realization or device use, offering even higher spin polarization than predicted for GaAs.

In Figure~\ref{spinor1}(C), we show the spectrum of $\cal DSP^{\mathrm{z}}$ for the graphane configuration (C) as is depicted in Figure~\ref{opt}(C). At the onset of absorption, which occurs at 3.30 eV, the injected electrons are 32\% spin-polarized, and the $\cal DSP^{\mathrm{z}}$  spectrum rises sharply. At photon energies greater than 3.30 eV, the  $\cal DSP^{\mathrm{z}}$  signal starts to decrease almost linearly, and at 3.80 eV the $\cal DSP^{\mathrm{z}}$ achieves a value of 17\%, at which point it drops vertically to zero. At photon energies greater than 3.80 eV and less than 3.30 eV, the  $\cal DSP^{\mathrm{z}}$  signal is zero. The overall shape of the spectrum $\cal DSP^{\mathrm{z}}$  displayed in  Figure~\ref{spinor1}(C) has the appearance of a right-trapezoid, and the excitation-energy window spans approximately 0.50 eV. We propose that the graphane configuration (C) is a favorable candidate for the experimental development of spintronic devices, because it provides a roughly 32\% spin polarization.  This value is comparable to the roughly -30\% spin polarization predicted for bulk Si by Nastos et al.~\cite{PhysRevB.76.205113}.  Our calculations indicate that the graphane configuration (C) has a direct gap at the $\Gamma$ point; therefore, the onset of  $\cal DSP^{\mathrm{z}}$  will be dominated by $\Gamma$-Point transitions.

Figure~\ref{spinor1}(D)  shows no $\mathcal{DSP}^{z}$ signal because the armchair configuration does not allow optical spin injection at any photon energy. All four graphane structures --chair, boat, zigzag, and armchair-- share the same CH stoichiometry with one carbon and one hydrogen atom per formula unit. This raises the question of why the armchair geometry forbids optical spin injection, whereas the chair, boat, and zigzag configurations permit it.

\subsection{Optical spin injection in fluorographene}
\begin{figure*}[htb!]
\begin{center}  
  \includegraphics[scale=0.95]{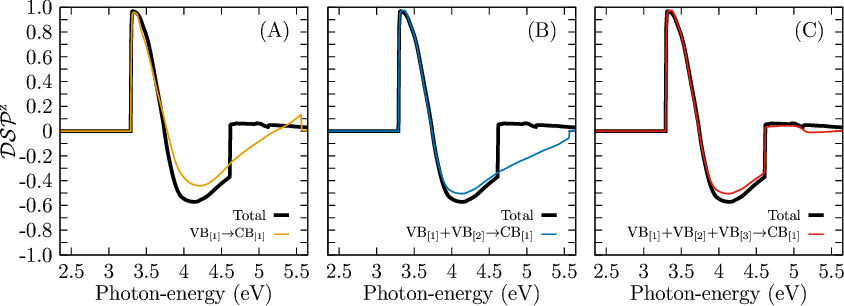}
  \caption{(Color online) Band-resolved decomposition of the ${\cal DSP}^{\mathrm{z}}$ for fluorographene configuration (C). The dominant optical transitions take place near the band edge and arise primarily from the upper valence bands into the lowest conduction band. The transition VB${[1]} \rightarrow$ CB${[1]}$ corresponds to the excitation from the first valence band to the first conduction band.
  VB[1] +VB${[2]}\rightarrow$ CB${[1]}$ corresponds to the excitation from the first and second valence bands to the first conduction band.}
\label{spin-band}
\end{center}    
\end{figure*}
The computed $\cal DSP^{\mathrm{z}}$ spectra for the four fluorographene configurations —(A) chair, (B) boat, (C) zigzag, and (D) armchair— are shown in Figure~\ref{spinor2}. 
In Figure~\ref{spinor2}(A), we display the $\cal DSP^{\mathrm{z}}$ spectrum for fluorographene configuration (A).  Inspection of Figure~\ref{spinor2}(A) reveals a narrow and intense  $\cal DSP^{\mathrm{z}}$ peak located at 2.74 eV, where the spin polarization sharply increases to  100\%. When the photon energy increases by just 0.08 eV, the $\cal DSP^{\mathrm{z}}$  value drops sharply and becomes zero at 2.82 eV. At this energy,  the  $\cal DSP^{\mathrm{z}}$ changes sign and reaches –10\%.
For photon energies above 2.82 eV, the $\cal DSP^{\mathrm{z}}$  value approaches zero.
The extremely narrow excitation window of only 0.08 eV suggests that this configuration may
have limited practical applicability, as efficient spintronic devices typically require a broader and more stable spin-polarization response.

In Figure~\ref{spinor2}(B), we show the $\cal DSP^{\mathrm{z}}$ spectrum for fluorographene in configuration (B). The onset of the  $\cal DSP^{\mathrm{z}}$ signal occurs at 3.05 eV and increases sharply, reaching a spin polarization of 74\%. As photon energy increases, the $\cal DSP^{\mathrm{z}}$ decreases linearly, changing sign and reaching, at a photon energy of 4.24 eV, a value of -11\% spin polarization. A second enhancement of the $\cal DSP^{\mathrm{z}}$ occurs at 4.26 eV,  reaching 88\%. With a slight increase in photon energy, the  $\cal DSP^{\mathrm{z}}$ signal drops rapidly to reach a value close to zero.    For photon energies above 4.80 eV, the signal is almost zero.   The first $\cal DSP^{\mathrm{z}}$  peak spans an energy interval from 3.05 to approximately 4.05 eV; this suggests that the fluorographene configuration (B)  may be suitable for experimental spintronic applications.

In Figure~\ref{spinor2}(C), we show the $\cal DSP^{\mathrm{z}}$ spectrum for fluorographene in configuration (C). A close analysis of the figure reveals that the onset of the 
$\cal DSP^{\mathrm{z}}$  signal occurs at 3.28 eV, where the signal increases abruptly, reaching 98\% spin-polarization. When the photon energy is increased, the  $\cal DSP^{\mathrm{z}}$  of electrons starts to decrease steadily until it crosses zero at 3.72 eV and becomes negative, reaching –58\% spin polarization. At photon energies greater than 4.13 eV, the $\cal DSP^{\mathrm{z}}$ signal begins to decline toward zero. For energies above 6.14 eV, the  $\cal DSP^{\mathrm{z}}$  remains zero. The excitation photon energy window associated with the main positive polarization peak spans 0.44 eV. The excitation-photon energy range that yields a positive and negative non-zero  $\cal DSP^{\mathrm{z}}$ spans 1.32 eV. Moreover,  a slight change in photon energy reverses the spin polarization. Based on our calculations, we suggest that the fluorographene configuration (C) may offer favorable conditions for experimental spintronics applications.

In Figure \ref{spinor2}(D), we show the $\mathcal{DSP}^{z}$ spectrum for fluorographene in configuration (D). The onset of the $\cal DSP^{\mathrm{z}}$ spectrum occurs at a photon energy of 4.23~eV, where a small negative $\mathcal{DSP}^{z}$  peak appears with a value of -{15\%}. There,  the $\mathcal{DSP}^{z}$  rises almost vertically and, with a slight increase of the photon energy, at around photon energy of 4.51~eV,  the $\mathcal{DSP}^{z}$  drops sharply to zero.  Beyond  4.91 eV photon-energy, the $\mathcal{DSP}^{z}$ spectrum is zero. The photon-energy window that allows spin-polarized injection spans 0.61 eV. The poor spin-polarization response in the fluorographene configuration (D) suggests that it is unsuitable for optospintronic applications.

Several recent works have demonstrated that enhancing SOC can increase the  ${\cal DSP}^{\mathrm{a}}$ of optically injected electrons in semiconductors. Bhat et al. showed that the DSP can reach ~78\% in GaSb and demonstrated that interband SOC plays a key role in achieving high ${\cal DSP}^{\mathrm{a}}$~\cite{PhysRevB.71.035209}. Moreover, Choi et al. reported that heavy metals with strong SOC can generate efficient optical spin-polarized electrons with circularly polarized light~\cite{https://doi.org/10.1002/adfm.202307753}. Earlier studies demonstrated that the hydrogenation and fluorination of graphene can significantly enhance the SOC~\cite{PhysRevLett.103.026804,Avsar2014}. 
Table~\ref{table_so2} lists the computed SOC values for the graphane and fluorographene configurations. The analysis of data in the Table reveals that the smallest SOC value belongs to the fluorographene configuration (B) (60~$\mu$eV), followed by the fluorographene configuration (C) (380~$\mu$eV), while the largest SOC value is for the graphane configuration (A) (2420~$\mu$eV).
Our results indicate that configurations with weaker SOC produce a high ${\cal DSP}^{\mathrm{z}}$ response over a broad photon-excitation window. Conversely, configurations with stronger SOC can reach ${\cal DSP}^{\mathrm{z}}$ values as high as 83–100\%, although only within a much narrower energy range. This behavior reflects the distinct mechanisms of SOC in functionalized graphene, where hybridization distortions critically determine the optical spin-injection efficiency. 
From our calculations, the ${\cal DSP}^{\mathrm{z}}$ spectrum for the fluorographene in configuration (C) exhibits one of the strongest  ${\cal DSP}^{\mathrm{z}}$ spin responses, with peak values of 98\% and -57\%
for positive and negative ${\cal DSP}^{\mathrm{z}}$, respectively. The photon-energy range with non-zero positive and negative ${\cal DSP}^{\mathrm{z}}$ spans 1.32 eV. The geometry of the fluorographene in configuration (C)  suggests that fluorine functionalization modifies the electronic states involved in optical transitions that promote electrons from valence-band to conduction-band states into a preferred spin state, thereby increasing spin polarization. A closer analysis of the data presented in Table~\ref{angulos} reveals that fluorographene configuration (C) has the smaller angle,  $\theta$[CCF],  with a value of 103.6$^{\circ}$ and the larger angle, $\theta$[CCC], with a value of 115.5$^{\circ}$. Fluorine adsorption increases the local sp$^3$ character of graphene, but configuration (C) does not achieve an ideal tetrahedral geometry. Ideal sp$^3$ geometry has angles near 109.5$^{\circ}$~\cite{CS9922100059}. In fluorographene configuration (C), the distorted angles indicate mixed sp$^2$–sp$^3$ hybridization. As a result, configuration (C) supports a stronger and wider ${\cal DSP}^{\mathrm{z}}$  response compared to other graphane and fluorographene configurations studied in this work.

\subsection{Band-resolved decomposition of ${\cal DSP}^{\mathrm{z}}$  spectra}
The calculation of the ${\cal DSP}^{\mathrm{z}}$ spectra is carried out using the sum-over-states formalism, which depends on the position matrix elements and the energy differences between bands. Within the band-resolved decomposition framework, we analyze and identify the contributions of different band-to-band transitions to the ${\cal DSP}^{\mathrm{z}}$ spectrum by summing only selected valence and conduction bands~\cite{PhysRevB.76.205113,ma13194300,PhysRevB.70.235110}. 
This approach allows us to analyze the individual contributions of different band transitions to the ${\cal DSP}^{\mathrm{z}}$ spectrum. Figure~\ref{spin-band} displays the band-by-band contributions to the ${\cal DSP}^{\mathrm{z}}$ response for selected transitions in fluorographene configuration (C).  In Figure~\ref{spin-band}(A, B, C, D), the black solid line represents the total ${\cal DSP}^{\mathrm{z}}$ spectrum obtained by summing over all valence and conduction band contributions.  In Figure~\ref{spin-band}(A), we show the specific contribution arising from transitions between the highest valence band and the lowest conduction band, VB${[1]} \rightarrow$ CB${[1]}$, is depicted by the yellow solid line. Interestingly, the complete positive peak of the ${\cal DSP}^{\mathrm{z}}$ spectrum and  negative peak in the ${\cal DSP}^{\mathrm{z}}$ spectrum are due to transitions from  the Y—$\Gamma$—X valley, particularly around to the $\Gamma$ point and the top of valence band and the lowest conduction band. We find that these transitions are responsible for most of the total ${\cal DSP}^{\mathrm{z}}$ response in fluorographene configuration (C). A closer inspection of the left panel in Figure~\ref{pdos} reveals that the top valence band is composed of F-p and C-p atomic states. Interestingly, most of the ${\cal DSP}^{\mathrm{z}}$ signal arises from transitions involving only the top valence bands and the lowest conduction bands. By chemical doping graphene with fluorine, we increase transitions from the top valence band to the lowest conduction band; as a result, the spin polarization increases.  Our results are in good agreement with Sipe et al., who reported that one way to increase the spin polarization of injected electrons is to use materials in which strain removes band degeneracy, allowing electrons to be excited from a single valence band~\cite{PhysRevB.71.035209}. Figure~\ref{spin-band}(B) presents the contribution to the ${\cal DSP}^{\mathrm{z}}$ spectrum originating from transitions between the two upper valence bands and the lowest conduction band, VB${[1]}$ + VB${[2]} \rightarrow$ CB$_{1}$, represented by the blue solid line. The inclusion of a second valence band results in an increase in the  ${\cal DSP}^{\mathrm{z}}$ spectrum (blue solid line)  to reach almost the total ${\cal DSP}^{\mathrm{z}}$ spectrum. Figure~\ref{spin-band}(C) displays the thin red curve for the VB${[1]}$ + VB${[2]}$ + VB${[3]} \rightarrow$ CB${1}$ contribution, When transitions from these three valence bands are included, the resulting spectrum nearly reproduces the total ${\cal DSP}^{\mathrm{z}}$ response.

From a chemistry standpoint, the higher electronegativity of fluorine modifies the sp$^3$ character in fluorographene. This modification of the hybridization, rather than the ionic or covalent nature of the bond, ultimately influences the resulting  ${\cal DSP}^{\mathrm{z}}$  reported in this work. Further studies are needed to investigate the influence of the bonding character on the  ${\cal DSP}^{\mathrm{z}}$  spectra.
\vspace{-0.5cm}
\section{Conclusions}
In summary, we computed the degree of spin polarization, ${\cal DSP}^{\mathrm{z}}$, of optically injected electrons generated by one-photon absorption in several graphane and fluorographene configurations.
Our findings show that the zigzag fluorographene configuration (C) exhibits a
${\cal DSP}^{\mathrm{z}}$ of 98\% for optically injected spin-polarized electrons
at the band edge. This response later undergoes a sign inversion, reaching a
negative value of -57\%. The excitation-photon energy range that yields a positive and negative non-zero  ${\cal DSP}^{\mathrm{z}}$  spans 1.32 eV. Moreover, a slight change in photon energy reverses the spin polarization. Based on our calculations, the fluorographene configuration (C) may offer favorable conditions for experimental spintronics applications. We note that the fluorographene configuration (C) exhibits the minimum  structural distortion relative to pristine graphene, graphane, and the other fluorographene configurations.   As a result, fluorographene configuration (C) exhibits the minimum sp$^2$–sp$^3$ hybridization compared to other graphane and fluorographene configurations studied in this work. Even though electrons are 100\% spin-polarized in the fluorographene configuration (A), this polarization occurs over a narrow photon-energy range, making this unfavorable for the development of spintronic devices.  Configurations (D) of graphene and fluorographene are entirely unfavorable for the development of spintronic devices. Graphane configurations (B) and (C) could also be employed to develop experimental spintronics. Regarding the SOC effect on the ${\cal DSP}^{\mathrm{z}}$  spectra. Our results indicate that the best  ${\cal DSP}^{\mathrm{z}}$ signal (largest value and wide window of the photon energy excitation) is exhibited by the structure that shows one of the lower SOC values. Interestingly, there is a correlation between SOC and the ${\cal DSP}^{\mathrm{z}}$ spectrum, but the results show that the structure with the largest SOC does not necessarily yield the best ${\cal DSP}^{\mathrm{z}}$ signal. The band-resolved decomposition analysis shows that almost the entire ${\cal DSP}^{\mathrm{z}}$ spectrum comes from transitions involving the top-valence band, composed of a mixture of the C-p and F-p states. This work demonstrates that graphene functionalization plays a significant role in optical spin-polarization response. As future work, two-dimensional materials offer a promising platform for studying optical spin orientation and excitonic effects, key ingredients for the next generation of optoelectronic and spintronic devices.

\section{Acknowledgments}
A. M. L.-M. thanks SECIHTI-M\'exico for the Ph.D. scholarship (1006569). We are also grateful to the computational chemistry laboratory for providing computational resources, \emph{ELBAKYAN}, and \emph{PAKAL} supercomputers of the  Polytechnic University of Tapachula. The authors thank the partial support of the PIFIP 2024 Project granted (UAdeO-22072025) by the Universidad Aut\'onoma de Occidente (UAdeO).

\section{Conflicts of Interest} The authors declare no conflict of interest.

\section{Funding} This research received partial funding under grant No. UAdeO-22072025.

\section{Author Contribution}
\vspace{-0.5cm}
A.M. L.-M. conceived the research idea, programmed code, performed the DFT simulations, generated the DSP spectra, carried out data analysis, wrote the manuscript, and contributed to visualization. C. C.-F. and A. L.-V. performed the DFT simulations, assisted in computational setup, structural modeling, and discussion of results. C. C.-Q. contributed to the interpretation of data and manuscript revision. J.L.C. and H.V.-B. conceived the research idea, programmed code, supervised the project, guided the theoretical methodology, wrote the manuscript, and contributed to the writing and scientific discussion. All authors reviewed the manuscript and approved the final version.
\bibliographystyle{unsrt}
\bibliography{bibliography.bib}
\newpage 
\typeout{get arXiv to do 4 passes: Label(s) may have changed. Rerun}
\end{document}